\documentstyle[preprint,cite,epsf,aps]{revtex}

\begin{document}
\tightenlines
\draft
\title{Renormalization Group Study of Chern-Simons Field
Coupled to Scalar Matter in a Modified BPHZ Subtraction Scheme\footnote{Copyright by The American Physical Society}}
\author{L. C. de Albuquerque, M. Gomes
and  A. J. da Silva\footnote{e-mails:{\it ajsilva,claudio,
mgomes@fma.if.usp.br}}}
\address{Instituto de F\'\i sica, USP\\
 C.P. 66318 - 05389-970, S\~ao\ Paulo - SP,  Brazil}
\date{\today}
\maketitle
\begin{abstract}

We apply a soft version of the BPHZ subtraction scheme
to the computation of two-loop corrections 
from an  Abelian Chern-Simons field coupled to (massive) scalar matter
with a $\lambda(\Phi^\dag\Phi)^2$ and $\nu(\Phi^\dag\Phi)^3$ self-interactions.
The two-loop renormalization group functions 
are calculated. We  compare our results with 
those in the literature.
\end{abstract}
 
\vspace{4cm}

\vfill
\newpage
\section{Introduction}
Field theories with the Chern-Simons (CS) term in $2+1$ dimensions
\cite{CS2,CS1} are among the best studied models in the past two
decades.  This is due not only to its potential applications but also
because of some subtle conceptual and technical aspects. Indeed, the
quantization of these theories raises interesting questions some of
which have not been answered satisfactorily up to now. Of particular
interest is the set up of a renormalization/regularization scheme
simple and reliable to provide a consistent framework for complex
calculations. Many proposals have appeared in the
literature\cite{CS1,Gaume,Asorey,Pimentel,Chen,Slavnov} each of them
presenting advantages as well disadvantages.  For instance,
Pauli-Villars regularization explicitly breaks parity, analytic
regularization is not gauge invariant, and Slavnov regularization
becomes rather intricate beyond one-loop calculations.

It is of course desirable that the regularization scheme
preserves as much of the models symmetries as possible.
In this respect, the popular dimensional regularization appears to be  suited 
for the task. However, the topological nature of the CS term introduces
extra complications. The Levi-Civita symbol
does not admit a simple extension to complex dimensions $D$ and
to overcome  this problem some modifications  have to be done\cite{Del}. 

If one insists in keeping the Levi-Civita symbol in three dimensions, other
approaches are possible. Firstly, there is the so called 
{\sl consistent dimensional regularization} \cite{Martin,Breit}
in which $\epsilon_{\mu\nu\lambda}$ is treated as essentially three
dimensional, but one has to introduce a Maxwell or Yang-Mills 
kinetic term as a supplementary regularization. In another proposal,
called {\sl dimensional reduction} \cite{Chen,Kaza,Gomes1,pinheiro}
the tensor algebra is done in three dimensions and afterwards the 
Feynman integrals are promoted to $D$ dimensions.  This method
may introduce
ambiguities in the {\sl finite} parts of the amplitudes 
and also in the divergent parts in high order corrections.

Now, it is possible to renormalize divergent integrals without introducing
an specific regularization. One of the most efficient
and rigorous method is the BPHZ subtraction scheme
which has been applied in a variety of situations\cite{BPHZ}.  One class of this method, the soft BPHZ schemes, is specially adequate for the study
of massless theories\cite{Gomes2}. 

In a soft BPHZ
scheme one introduces subtraction operators in the external
momenta and in the masses of the theory.
In order to have a better control of the infrared divergences, some
of the subtractions have zero mass whereas  others have
mass equal to $\mu$ (the renormalization scale).
In $D=4$ this scheme is in many circumstances  equivalent to the dimensional 
regularization with a minimal subtraction prescription,
and leads to a mass-independent renormalization group equation \cite{RGE}.

In this work we use a soft BPHZ scheme in an Abelian CS theory coupled
with scalar matter to compute two-loop renormalization group
functions. Since this method do not involve analytic continuation in
the space-time dimension (i.e., we stay in the physical dimension
$D=3$), we evade the problems aforementioned. Although we will be
dealing mainly with massive particles the scheme allows, for
non-exceptional momenta, a smooth zero mass limit if all
super-renormalizable interactions are deleted.  The paper is organized
as follows: In section II we introduce the model, generalize the soft
BPHZ scheme to the CS theory with scalar matter, and show that the
soft BPHZ scheme respects the Ward identity.  In Section III we
compute the renormalization group functions up to two-loops. Finally,
we draw some conclusions and comments.  Our results for the
renormalization group functions extend the ones recently computed
through the use of the {\it dimensional reduction} \cite{pinheiro}, in
the sense that the results agree whenever a comparison is possible.
The mentioned agreement indicates that
{\it dimensional reduction} is a consistent scheme, at least up two-loops,
as far as renormalization group functions are concerned.
However, differently from \cite{pinheiro} the model that we study includes the most general,
renormalizable and $U(1)$ invariant, self-interaction of the scalar
particles.

\section{The Model and the Soft BPHZ Scheme}
\label{S1}

The Lagrangian describing the model reads (the metric has signature $(+,-,-)$ and $\epsilon^{012}=1$)

\begin{eqnarray} \label{1} 
{\cal L}=& &(D_\mu\Phi)^\dagger D^\mu\Phi -m^2\Phi^\dagger\Phi
 -\frac{\lambda}{6}\bigl(\Phi^\dagger\Phi\bigr)^2
\nonumber\\
& &
-\frac{\nu}{36}\bigl(\Phi^\dagger\Phi\bigr)^3
+\frac{1}{4}\epsilon^{\mu\nu\lambda}\,F_{\mu\nu}\,A_\lambda 
-\frac{1}{2\xi}(\partial\cdot A)^2,
\end{eqnarray}

\noindent
where $D_\mu=\partial_\mu -ie A_\mu$ is the covariant derivative and 
 $F_{\mu\nu}=\partial_\mu\,A_\nu-\partial_\nu\,A_\mu$. 
The canonical operator
dimensions are as follows (in unity of mass: $[m]=1$):
$[\Phi]=1/2$, $[A_\mu]=[\lambda]=[\xi]=1$,  $[\nu]=[e]=0$.
The ultraviolet degree of superficial divergence of a
graph $\Gamma$ is 

\begin{equation}
d(\Gamma)=3-V_{\phi^4}-\frac{1}{2}N_\phi-N_A,\label{2}
\end{equation}
where $V_{\phi^4}$ is the number of insertions of the
super-renormalizable vertex $\lambda\bigl(\Phi^\dagger\Phi\bigr)^2$,
and $N_\phi$ ($N_A$) is the number of scalar (vector) external lines
in $\Gamma$.  Nonetheless, we must stress that, due to the
contractions of indexes, this degree is lowered by the number of
external trilinear vertices having attached one internal CS line.  In this
work we will use the Landau gauge, formally obtained by letting $\xi
\rightarrow 0$.

In a soft BPHZ approach the divergent  quantum amplitudes, 
described by Feynman integrals in perturbation theory, are made finite 
by the application of subtraction operators $\tau^{d(\gamma)}$  arranged in accord with the {\it forest formula}\cite{BPHZ}. 

The subtraction operator $\tau^n$ has degree $n$ in the derivatives
and uses derivatives with respect to $m^2$ (each of them having degree
2) if the integrand depends only on the square of this mass. All
infrared finite subtractions are made at $m^2=0$ but in the would be
infrared divergent terms (possibly occurring in the last subtractions)
we make $m^2=\mu^2$. As a rule, we make the minimum number of
subtractions necessary to render finite a Feynman amplitude.
Explicitly, the first three subtraction operators are defined as

\begin{eqnarray}\label{3} 
& & \tau^0\,{\cal G}(p_i,m)={\cal G}(0,\mu),\nonumber\\ 
& & \tau^1\,{\cal G}(p_i,m)={\cal G}(0,0)+
p^\mu_i\frac{\partial {\cal G}}{\partial\,p^\mu_i}
\Bigl |_{p_i=0,m=\mu}\Bigr .\, ,\\ 
& & \tau^2\,{\cal G}(p_i,m)={\cal G}(0,0)+
p^\mu_i\frac{\partial{\cal G}}{\partial\,p^\mu_i}
\Bigl |_{p_i=0,m=0}+\frac{1}{2}p^\mu_i p^\nu_j\frac{\partial^2{\cal G}}
{\partial p^\mu_i\partial p^\nu_j}\Bigl |_{p_i=0,m=\mu}
+m^2\frac{\partial{\cal G}}{\partial m^2}\Bigl |_{p_i=0,m=\mu}\Bigr.\Bigr.
\Bigr.\,,\nonumber\\
& & \tau^3\,{\cal G}(p_i,m)={\cal G}(0,0)+
p^\mu_i\frac{\partial{\cal G}}{\partial\,p^\mu_i}
\Bigl |_{p_i=0,m=0}+\frac{1}{2}p^\mu_i p^\nu_j\frac{\partial^2{\cal G}}
{\partial p^\mu_i\partial p^\nu_j}\Bigl |_{p_i=0,m=0}
+m^2\frac{\partial{\cal G}}{\partial m^2}\Bigl |_{p_i=0,m=0}\Bigr.\Bigr.
\Bigr.\nonumber\\
& &\qquad\qquad
+\frac{1}{6}p^\mu_i p^\nu_j p^\lambda_k\frac{\partial^3{\cal
  G}}
{\partial p^\mu_i\partial p^\nu_j \partial p^\lambda_k}
\Bigl |_{p_i=0,m=\mu}+m^2 p^\mu_i\frac{\partial^2{\cal
  G}}
{\partial p^\mu_i\partial m^2 }\Bigl |_{p_i=0,m=\mu}\Bigr.\Bigr.\,.\nonumber
\end{eqnarray}

\noindent
 It is apparent that our scheme does
not introduce infrared divergences in graphs having at least one
internal scalar line. For ultraviolet divergent graphs without
internal scalar lines, a special care has to be taken.  To be precise,
for ultraviolet divergent graphs without internal scalar lines, we will
use the regularized CS field propagator

\begin{equation}
\Delta_{\mu\nu}(p)= \epsilon_{\mu\nu\rho} \frac{p^\rho}
{p^2-M^2+i\epsilon}\,\, ,\label{31}
\end{equation}

\noindent
where $M$ is a regularization mass which, in the subtraction terms,
is to be treated in the same way as $m$ and has to be put equal to zero
after all the subtractions are performed.
 
To see how the method works, in the following we will compute
the vacuum polarization at one-loop in the theory specified by
Eq. (\ref{1}). The vacuum polarization is given by the sum of two
terms, namely:

\begin{eqnarray}
\label{4B}
& & \pi^{(A)}_{\rho\lambda}=2ie^2 g_{\rho\lambda}\int
[dq]\,\Delta(q),\\
& & \pi^{(B)}_{\rho\lambda}(p)=-e^2\int [dq]\,(2q+p)_\rho
(2q+p)_\lambda\,\Delta(q)\,\Delta(q+p)\equiv - e^2\int [dq]\, I_{\rho\lambda}(p,q;m),
\label{4B1}
\end{eqnarray}

\noindent
where $\Delta (k)=i/\bigl(k^2-m^2+i\epsilon\bigr)$ is the scalar
propagator, and $[dq]=d^3 q/(2\pi)^3$. Both
$\pi^{(A)}_{\rho\lambda}$ and $\pi^{(B)}_{\rho\lambda}$ 
are linearly divergents. For the renormalized quantities we obtain

\begin{eqnarray}
 \pi^{(A)}_{\rho\lambda}\Big\vert_R&=&2ie^2 g_{\rho\lambda}\int [dq] (1-\tau^1)\,\Delta(q)
=-2e^2 g_{\rho\lambda}\int
[dq]\,\Biggl[\frac{1}{q^2-m^2}-\frac{1}{q^2}
\Biggr]\, ,\label{4C1}\\
 \pi^{(B)}_{\rho\lambda}\Bigl |_R(p)&=&-e^2\int [dq](1-\tau^1)\,I_{\rho\lambda}(p,q;m)\nonumber\\
& =&e^2 \int[dq]\Biggl[
\frac{(2q+p)_\rho(2q+p)_\lambda}{(q^2-m^2)\bigl((q+p)^2-m^2\bigr)}
-4\frac{q_\rho q_\lambda}{(q^2)^2}- p^\alpha\frac{\partial I_{\rho\lambda}}{\partial p^\alpha}\Biggl |_{m=\mu,p=0}\Biggr.\Biggr]\Bigr.\, .\label{4C2}
\end{eqnarray}

\noindent
From a practical standpoint, 
the  finite integrals in Eqs. (\ref{4C1}) and  (\ref{4C2})
may be most easily computed using an intermediate regularization
(cutoff, dimensional, Pauli-Villars, etc), the final result being of course 
independent of the regularization employed. Note  that 
the last term of the Eq. (\ref{4C2}) vanishes upon symmetric integration.
We obtain,

\begin{eqnarray}
& & \pi^{(A)}_{\rho\lambda}\Big\vert_R=-i\frac{e^2\,m}{2\pi}\, 
g_{\rho\lambda},\\
& & \pi^{(B)}_{\rho\lambda}(p)\Biggl|_R=i\frac{e^2}{8\pi}\Biggl[\,2m\,\biggl(
g_{\rho\lambda}+\frac{p_\rho p_\lambda}{p^2}\biggr)
-\frac{p^2-4m^2}{\sqrt{p^2}}\sinh^{-1}\Bigl(\frac{\sqrt{p^2}}
{\sqrt{4m^2-p^2}}\Bigr)\,{\cal T}_{\rho\lambda}\Biggr]\Bigr.,\label{4C3}
\end{eqnarray}

\noindent
where ${\cal T}_{\rho\lambda}=g_{\rho\lambda}-p_\rho p_\lambda/p^2$.
Using that $\pi_{\rho\lambda}(p)=\pi^{(A)}_{\rho\lambda}+
\pi^{(B)}_{\rho\lambda}(p)$, we finally get (note that
$\pi(p^2\rightarrow0)=0$ as it should)

\begin{equation}\label{4D}
\pi_{\rho\lambda}(p)\Big\vert_R=\pi(p){\cal T}_{\rho\lambda}=-i\frac{e^2}{8\pi}\Biggl[\,2m+\frac{p^2-4m^2}{\sqrt{p^2}}
\sinh^{-1}\Bigl(\frac{\sqrt{p^2}}{\sqrt{4m^2-p^2}}\Bigr)\Biggr] 
{\cal T}_{\rho\lambda}.
\end{equation}

The Lagrangian density in Eq. (\ref{1}) is invariant under global $U(1)$ 
transformations. At the classical level, this demands the conservation 
of the current
$J_\mu(x)=
ie[\Phi^\dagger\stackrel{\leftrightarrow}{\partial}_\mu\Phi] +2e^2
\Phi^\dagger\Phi A_\mu$.  At the quantum level  this current will be
 quantized with normal product of minimum degree, i. e.  two.  The corresponding
Ward identity for the Green functions reads

\begin{eqnarray}\label{9}
& &\partial_\mu \langle\,T\,N_2[J^\mu](x)\,X_{\{\nu_j\}}
(x_k,y_j)\,\rangle= \langle\,T\,N_3[\partial^\mu J_\mu](x)\,X_{\{\nu_j\}}(x_k,y_j)\,\rangle
 \nonumber\\
& &\qquad=e\,\Biggl[\,\sum_{j=1}^{N}\delta (x-x_j)\,
-\sum_{j=N+1}^{2N}\delta (x-x_j)\Biggr]\,
\langle\,T\,X_{\{\nu_j\}}
(x_k,y_j)\,\rangle \, ,
\end{eqnarray}

\noindent
where $X_{\{\nu_j\}}
(x_k,y_j) =\prod_{k=1}^{N}\Phi(x_k)\,{\prod}_{i=N+1}^{2N}
\Phi^\dagger (x_i)\,\prod_{l=1}^L A_{\nu_l}(y_l)$, follows from the
use of the normal product algorithm which turns out to be valid in
our scheme. In the first step Lowenstein's differentiation rule \cite{BPHZ}
was used to get the partial derivative inside the
normal product (which can be done if one increases the degree of the
normal product by one unity); in the second step one uses the
equations of motion for the bilinears $\Phi^{\dagger}\Box \Phi$ and $\Phi\Box 
\Phi^{\dagger}$, which in our case have the classical form. 

The Ward identity for the proper vertex functions has the same form as
in Eq. (\ref{9}) except for a minus sign on its right hand side.
Denoting by $\Gamma^{(2N,L)}(p_1\ldots,p_{2N};k_1,\ldots k_L)$ the proper vertex function of 
equal numbers ($N$) of  $\Phi$ and $\Phi^\dagger$ fields and $L$ gauge
fields, one can verify that, in momentum space

\begin{eqnarray}
q^\mu\, \Gamma_{\mu}^{(2,1)}(p,p^\prime;q)&=& - e [\Gamma^{(2)}(p^\prime)-
\Gamma^{(2)}(p)],\\
q^\mu\, \Gamma_{\mu\nu}^{(2,2)}(p,p^\prime;q,k)&=& - e [\Gamma_{\nu}^{(2,1)}
(p+q,p^\prime;k)-
\Gamma_{\nu}^{(2,1)}(p,p^\prime-q;k)].
\end{eqnarray}
Here and in the following  we adopt the simplified notation
$\Gamma^{(n)}\equiv \Gamma^{(n,0)}$.

\section{Renormalization Group Functions}
\label{S2}

The renormalized vertex functions introduced in the previous section satisfy 
the renormalization group equation

\begin{equation}
\label{12}
\Biggl[\,\mu\frac{\partial}{\partial\mu}+
\beta_{m^2}\frac{\partial}{\partial m^2
}+\beta_\lambda\frac{\partial}{\partial\lambda}+\beta_\nu
\frac{\partial}{\partial\nu}+
\frac{1}{2}\beta_{e^2}\frac{\partial}{\partial
e^2}-N\gamma_{\Phi} -L\gamma_A
\,\Biggr]\,\Gamma^{(N,L)}=0,
\end{equation}

\noindent
where $\beta_{m^2}$, $\beta_\lambda$, $\beta_\nu$, $\beta_{e^2}$,
$\gamma_{\Phi}$ and $\gamma_A$ are power series in the coupling
constants $e^2,\,\nu$, and $\lambda$.  We note that $\beta_{e^2}= e^2
\gamma_A=0$ as a consequence of the Coleman-Hill
theorem\cite{Coleman}.  

We shall fix
now the other functions. We begin by proving that they  do  not have
one-loop contributions. Indeed, in the computation of the
renormalization group functions the relevant contributions come
through the $\mu$ dependence of the subtraction terms.  As mentioned
before, these subtractions are those which are potentially infrared
logarithmically divergent. We shall now examine these possible
contributions.  We will use a graphical notation in which each diagram
represents a set of Feynman graphs differing by the orientation of the
external lines. Moreover, to facilitate the discussion, one may use an
auxiliary regularization so that each subtraction can be analyzed
individually.

\begin{figure}
\centerline{ \epsfbox{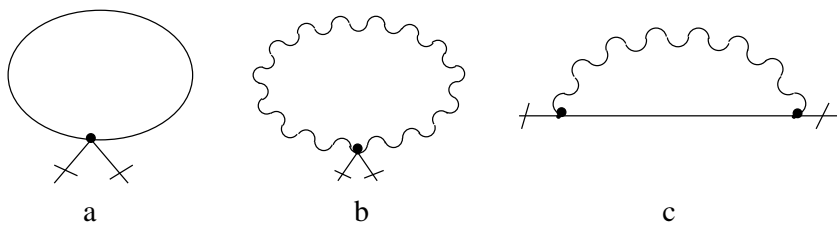}}
\caption{{\it $\Gamma^{(2)}$ at one loop.}} \label{fig1}
\end{figure}

\begin{figure}
\centerline{ \epsfbox{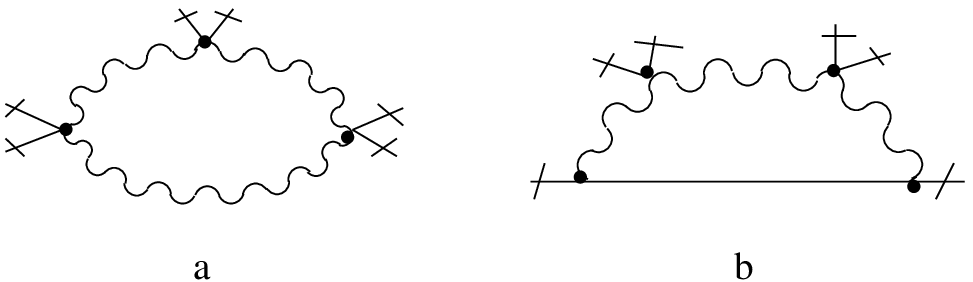}} \caption{{\it Divergent contributions
    to $\Gamma^{(6)}$ at one loop.}} 
\label{fig2}
\end{figure}

\begin{figure}
\centerline{ \epsfbox{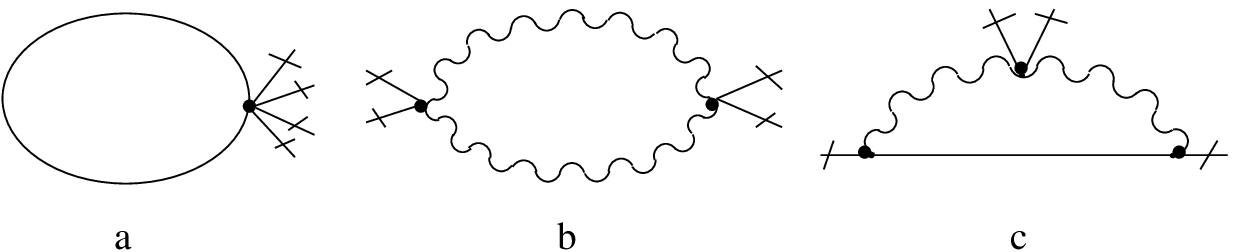}} \caption{{\it Divergent contributions
    to $\Gamma^{(4)}$ at one loop.}} \label{fig3}
\end{figure}

The divergent graphs contributing to $\Gamma^{(2)}$ at one-loop are
shown in Fig. \ref{fig1}.  Diagram \ref{fig1}a is linearly divergent
but, as it does not depend on the external momentum,  it does not lead to a
logarithmic term.  The graphs in
\ref{fig1}b and \ref{fig1}c vanish upon contraction of indexes. More
generally, the subtractions for any one-loop graph contributing to
$\Gamma^{(N)}$, with an odd number of CS lines, vanishes in the Landau
gauge.  This can be used to eliminate the possible contributions to
$\Gamma^{(6)}$ coming from Fig. \ref{fig2}.  Finally, Fig. \ref{fig3} shows 
divergent contributions
to $\Gamma^{(4)}$ at one-loop.  Graph \ref{fig3}a is linearly
divergent, with no logarithmic corrections. Graph \ref{fig3}b is
linearly divergent.  Using $\tau^1_M$ in Eq. (\ref{3}), and, in accord
with (\ref{31}), taking the limit $M\rightarrow0$ at the end, we found
that it does not depend on $\mu$. Diagram \ref{fig3}c is actually
finite due to the observations made after Eq. (\ref{2}). 
A similar analysis can be done for
$\Gamma^{(2,1)}$ and $\Gamma^{(2,2)}$ leading to the result mentioned
in the beginning of this section.

Let ${\Sigma}(p)$ be the self-energy function
defined by $\Gamma^{(2)} (p)=i\bigl[p^2-m^2-{\Sigma}
(p)\bigr]$. We have

\begin{equation}\label{14}
\Biggl[\,\mu\frac{\partial}{\partial\mu}+
\beta_{m^2}\frac{\partial}{\partial m^2
}+\beta_\lambda\frac{\partial}{\partial\lambda}+\beta_\nu
\frac{\partial}{\partial\nu}-2\gamma_{\Phi}\,\Biggr]\,{\Sigma} (p) 
+\beta_{m^2}+
2\gamma_{\Phi} (p^2-m^2)=0. 
\end{equation}
 
It is then easily verified that

\begin{eqnarray}
& &\mu\frac{\partial}{\partial\mu}\,{\Sigma}^{[1]}(p) 
+\beta_{m^2}^{[1]}+
2\gamma_{\Phi}^{[1]}(p^2-m^2)=0,\label{15}\\
& &\gamma_{\Phi}^{[1]}=-\frac{1}{4}\frac{p_\mu p_\nu}{p^2}
\frac{\partial^2}{\partial p^\mu\partial p^\nu}
\,\Biggl(\mu\frac{\partial}{\partial\mu}\,{\Sigma}^{[1]}\Biggr)
\Bigg\vert_{p=0},\label{16}\\
& & \beta_{m^2}^{[1]}=2 m^2\gamma_{\Phi}^{[1]}-
\mu\frac{\partial}{\partial\mu}\,{\Sigma}^{[1]}\Bigg\vert_{p=0}, \label{17}
\end{eqnarray}

\noindent
where the superscript in parenthesis designates the order of the corresponding
quantity in the loop expansion. From Eqs. (\ref{16}) and (\ref{17}) we obtain
that $\gamma_{\Phi}^{[1]}=\beta_{m^2}^{[1]}=0$, as there is no $\mu$
dependence at one-loop. From Eq. (\ref{12}) follows then that
$\beta_\nu^{[1]}=\beta_\lambda^{[1]}=0$.

Let us now proceed to the two-loop calculation of the renormalization
group functions. It can be easily verified that $\gamma_{\Phi}^{[2]}$
and $\beta_{m^2}^{[2]}$ are given by Eqs. (\ref{16}) and (\ref{17}),
with the superscript [1] replaced by [2].  Besides that, using the
one-loop results, and writing the renormalization group equations for
$\Gamma^{(6)}$ and $\Gamma^{(4)}$ in a loop expansion, we obtain at
two-loops the following relations:

\begin{eqnarray}
& &\beta_{\nu}^{[2]}=
6\nu\gamma^{[2]}_{\Phi}+i\mu\frac{\partial}{\partial\mu}\Gamma^{(6)[2]}(0)
\label{18}\\
& &\beta_{\lambda}^{[2]}=4\lambda\gamma^{[2]}_{\Phi}+i\mu\frac{\partial}
{\partial\mu}\Gamma^{(4)[2]}(0).\label{19} 
\end{eqnarray}

\noindent
We have used the analytical continuation in the number of dimensions
as intermediate regularization. It turns out 
that all integrals 
needed for  the computations may be expressed in terms of
($n$-dimensional Euclidean space, with
$\{dq\}=\sigma^{\epsilon}\,d^n q/(2\pi)^n$,
where $\sigma$ is a mass scale, and $\epsilon=3-n$)

\begin{eqnarray}
& &I_1^{(n)}(M_a,M_b,M_c) \equiv \int \{dk\}\{ dq\}
\frac{1}{(k^2+M_a^2)(q^2+M_b^2)
\Bigl((k+q)^2+M_c^2\Bigr)}\nonumber\\
& &\qquad\qquad=\frac{1}{32\pi^2}\,\Biggl[\,\frac{1}{\epsilon}-\gamma+1
-\ln\Biggl(\frac{(M_a+M_b+M_c)^2}{4\pi\sigma^2}\Biggr)\,\Biggr].\label{22}\\
& &I_{2}^{(n)}(M_a,M_b)\equiv \int \{dk\}\{ dq\}\frac{1}{(k^2+M_a^2)
\Bigl((k+q)^2+M_b^2\Bigr)}=\frac{1}{16\pi^2}\,M_a\,M_b.\label{23}
\end{eqnarray}

At two-loops, the nonvanishing contributions to $\gamma_{\Phi}$ come
from the three (quadratically divergent) diagrams in Fig. \ref{fig4}.
The diagram \ref{fig4}a may be written as

\begin{equation}\label{20}
{\Sigma}^{[2]}_A(p)=
-2ie^4\int[dk] [dq]\,\Delta(k+q-p)\,\Delta^{\mu\nu}(k)\,
\Delta_{\mu\nu}(q)\Bigg\vert_{SBPHZ},
\end{equation}

This diagram has three divergent subgraphs and eight forests, but the
relevant term comes from the subtraction associated to the diagram as
a whole. An application of the forest formula leads to the following
contribution to $\gamma^{[2]}_{\Phi}$

\begin{eqnarray}\label{21}
& &\gamma_{\Phi \, A}^{[2]}=-\frac{e^4}{48\pi^2}\,.
\end{eqnarray}

\noindent 
The second graph, Fig. \ref{fig4}b, corresponds to the unsubtracted
integral

\begin{eqnarray}\label{24}
\Sigma^{[2]}_B(p)& &=ie^4\int[dk] [dq]\,(2p+k)^\mu (2p+2k+q)^\alpha
(2p+k+2q)^\nu (2p+q)^\beta\,\Delta(p+k)\nonumber\\
& &\qquad\times\Delta(p+k+q)\,
\Delta(p+q)\,\Delta_{\mu\nu}(k)\,\Delta_{\alpha\beta}(q)\Bigg\vert_{SBPHZ}\, .
\end{eqnarray}

\noindent
In this case there is no divergent subgraph and
the calculation of $\gamma^{[2]}_{\Phi\, B}$
is reduced to just the contribution from the forest corresponding to
the graph as a whole. We then get

\begin{equation}\label{25}
\gamma_{\Phi\, B}^{[2]}=-\frac{e^4}{12\pi^2}\,\, .
\end{equation}

\noindent 
The graph in Fig. \ref{fig4}c is written as

\begin{eqnarray}\label{26}
\Sigma^{[2]}_C(p)& &=ie^4\int[dk] [dq]\,(2p-k)^\mu (2q-k)^\alpha
(2q-k)^\beta (2p-k)^\nu\,\Delta(k-p)\nonumber\\
& &\qquad\times\Delta(q)\,\Delta(q-k)
\Delta_{\mu\alpha}(k)\,\Delta_{\beta\nu}(k)\Bigg\vert_{SBPHZ} .
\end{eqnarray}

\noindent
The calculation now is more involved and details will provided in the
Appendix. That analysis produces the result

\begin{equation}\label{27}
\gamma_{\Phi\, C}^{[2]}=-\frac{e^4}{24\pi^2}.
\end{equation}

Adding the results  in Eqs. (\ref{21}),
(\ref{25}) and (\ref{27}), we obtain
 
\begin{equation}\label{28}
\gamma_{\Phi}^{[2]}=-\frac{7 e^4}{48\pi^2} \, ,
\end{equation}

\noindent
which is in accord with the result obtained in \cite{pinheiro} using
{\it dimensional reduction}.

The diagrams that contribute with a logarithmic correction to
$\Gamma^{(6)[2]}$ are shown in Fig. \ref{fig5}; each one of them is
superficially logarithmically divergent. Concerning these graphs we
make the following comments.  Diagrams \ref{fig5}a and \ref{fig5}d do
not have divergent subgraphs. The other graphs, \ref{fig5}b-c and
\ref{fig5}e-f have just one divergent subgraph. However, in all cases
the contribution coming from the subgraph vanishes. For the graph
\ref{fig5}b this happens after the contraction of indexes whereas for the
graphs \ref{fig5}e and \ref{fig5}f this results from a cancellation between
different forests; also, the possible contribution arising from the
subgraph of \ref{fig5}c is just the graph \ref{fig3}b whose $\mu$ dependent
subtractions vanish under symmetric integration.  The conclusion is
that in each case one has to compute only the contribution of the
forest containing just the graph as a whole. The results of these
calculations are summarized in table \ref{tb1}. Summing those contributions
and using Eq. (\ref{18}) we obtain

\begin{equation}\label{35} 
\beta_{\nu}^{[2]}=\frac{7}{24\pi^2}\,\nu^2-\frac{5}{\pi^2}\,
e^4\nu+\frac{72}{\pi^2}\,e^8.
\end{equation}

This result coincides with that of \cite{pinheiro}. To obtain
$\beta^{[2]}_\lambda$ we need to compute logarithmic contributions
arising from $\Gamma^{(4)[2]}$. The relevant graphs are listed in Fig.
\ref{fig6}.  The graph  \ref{fig6}a has $d(\Gamma)=0$, and does not
contain a divergent subdiagram. A straightforward calculation gives

\begin{equation}\label{36} 
\Gamma_A^{(4)[2]}=-\frac{i}{8\pi^2}\,\lambda\nu\,\,\ln\left
(\frac{m}{\mu}\right).
\end{equation}

\noindent 
The diagram \ref{fig6}b also has  $d(\Gamma)=0$, and no divergent
subdiagram. It leads to 

\begin{equation}\label{37} 
\Gamma_B^{(4)[2]}=\frac{i}{\pi^2}\,\lambda e^4\,\,\ln\left(\frac{m}{\mu}\right).
\end{equation}

\noindent The diagram  \ref{fig6}c has the diagram \ref{fig3}b as a divergent 
subdiagram.  We already know that this subgraph does not contribute.
The calculation of the the forest containing just the overall diagram
gives

\begin{equation}\label{38} 
\Gamma_C^{(4)[2]}=\frac{i}{2\pi^2}\,\lambda e^4\,\,\ln\Bigl
(\frac{m}{\mu}\Bigr).
\end{equation}

Taking the above results and using 
Eq. (\ref{19}) we obtain

\begin{equation}\label{39} 
\beta_{\lambda}^{[2]}=\frac{1}{8\pi^2}\lambda\nu
-\frac{25}{12\pi^2}\,e^4\lambda.
\end{equation}

We turn now to the computation of $\beta_{m^2}^{[2]}$.  According to Eq.
(\ref{17}), we still need to compute the graphs shown in Fig.
\ref{fig7}. The diagram \ref{fig7}a is the same as that in Fig.
\ref{fig4}a, now with $p=0$.  There is a cancellation between forests.
After some manipulations, we obtain

\begin{equation}\label{40} 
{\Sigma}_A^{[2]}(0)=-\frac{1}{8\pi^2}\,m^2 e^4
\,\ln\Bigl(\frac{m}{\mu}\Bigr).
\end{equation}

\noindent The diagram in \ref{fig7}b has $d(\Gamma)=2$, and three 
divergent subgraphs. However, due to cancellations the end result is

\begin{equation}\label{41} 
{\Sigma}_B^{[2]}(0)=-\frac{1}{4\pi^2}\,m^2 e^4
\,\ln\Bigl(\frac{m}{\mu}\Bigr) \, \, .
\end{equation}

\noindent Finally, graph \ref{fig7}c has $d(\Gamma)=0$ and no divergent
subdiagram. The result is

\begin{equation}\label{42} 
{\Sigma}_C^{[2]}(0)=\frac{1}{32\pi^2}\,\lambda^2\,
\ln\Bigl(\frac{m}{\mu}\Bigr)\, \, .
\end{equation}

Collecting the partial results, ${\Sigma}^{[2]} (0)=
{\Sigma}_A^{[2]}(0)+{\Sigma}_B^{[2]}(0)+{\Sigma}_C^{[2]}(0)$
from Eqs. (\ref{40})-(\ref{42}),
and using  Eq. (\ref{17}) with $n=2$, we obtain

\begin{equation} \label{43} 
\beta_{m^2}^{[2]}=\frac{1}{32\pi^2}\,\lambda^2
-\frac{2}{3\pi^2}\,m^2\,e^4\, \, .
\end{equation}

\noindent
For $\lambda=0$ this agrees with the anomalous dimension of the composite
operator $\Phi^\dagger\Phi$ as computed in \cite{pinheiro}, as it should.

To discuss the fixed points structure of the model we introduce a dimensionless
coupling by $\lambda \equiv \hat \lambda\mu$. Up to two-loops the {\it beta}
functions are given by Eq. (\ref{35}), 

\begin{equation}
\label{44}
\beta_{\hat\lambda}=-\hat \lambda+\frac{1}{8\pi^2}\hat\lambda\nu
-\frac{25}{12\pi^2}\,e^4\hat \lambda,
\end{equation}

\noindent
and by Eq. (\ref{43}) with $\lambda$ replaced by $\mu\hat \lambda$.
If $m$ and $\hat \lambda$ are zero there are no induction of
$\Phi^\dagger\Phi$ and $(\Phi^\dagger\Phi)^2$ counterterms as
$\beta_{\hat\lambda}$ and $\beta_{m^2}$ both vanish. The case $e=0$
has been analyzed in the literature \cite{Aragao} unveiling an
interesting tricritical behavior.  Near the trivial fixed point, for
small momenta the running couplings $\hat \lambda_{ef}$ and $\nu_{ef}$
are driven away and approach the origin, respectively. For large
momenta the opposite happens. As mentioned in \cite{pinheiro}, for
$e\not =0$ there are no other fixed points since, $\beta_\nu$ never
vanishes in the perturbative region. Besides, the behavior of $\hat 
\lambda_{ef}$ near the origin is not sensible to the introduction of $e$.

\section{Conclusions}
\label{Conc}

In this paper we have shown that it is possible to define a
consistent, gauge invariant subtraction scheme with a soft behavior in
the infrared regime (soft BPHZ) for an Abelian Chern-Simons theory
coupled to a massive scalar matter with  $\lambda(\Phi^\dag\Phi)^2$ and
$\nu(\Phi^\dag\Phi)^3$ self-interactions.  Within the soft BPHZ
approach we circumvent the problems associated with the analytical
continuation of the Levi-Civita tensor.  Hence, there is no need to
deal with a complicated {\it consistent dimensional regularization}
and neither it is necessary to introduce a Maxwell term. However,
there is a price for this simplification, since we have to deal with
various forests with the consequent increase in the number of Feynman
integrals. In the process of calculation, however, we have found that
these integrals can be reduced to a few primitive ones. In all cases
that we studied we explicitly verified the finiteness of the subtracted
integrals.

We have done a two-loop calculation of the renormalization group functions.
Analogous models were studied in \cite{Chen,Kaza,pinheiro,Hoso}.
Our $\gamma_{\Phi}^{[2]}$ agrees with the one computed in \cite{Chen}
(Abelian case). We found only a qualitative agreement with
\cite{Kaza} but a more close comparison seems infeasible
due to the lack of details in \cite{Kaza}.

The comparison with the renormalization group functions computed with
the {\it consistent dimensional regularization} scheme in \cite{Hoso}
is more difficult. Indeed, the RG functions computed in \cite{Hoso}
contain divergent contributions in the pure CS limit (no Maxwell term)
that are absent in our approach.  However there is some partial
agreement between our results and the finite parts of
$\beta_\lambda^{[2]}$, $\beta_{m^2}^{[2]}$, and $\beta^{[2]}_{\nu}$ of
\cite{Hoso} (some coefficients of the expansion of these functions are
identical to ours).  Our result for $\gamma_{\Phi}^{[2]}$ is  entirely
different from that in \cite{Hoso} as those authors claim that
$\gamma_{\Phi}^{[2]}=0$. The discrepancy could in principle be
attributed to the use of different renormalization schemes. The use of
an extra regularization represented by a Maxwell term for the $A^\mu$
field brings additional complications in their proposal as some of the
coefficients of the renormalization group functions become singular as
the regularization is removed.  A more careful analysis of this method
is still lacking.

Although our main interest resides in the pure CS model, we would like
to make a few comments on the possible changes in our scheme if a
Maxwell term $-\frac{a}4 F^{\mu\nu} F_{\mu\nu}$ is added to Eq. (\ref{1}). The
propagator for the gauge field is then modified to

\begin{equation}
\Delta_{\mu\nu}(p) = \epsilon_{\mu\nu\rho}\frac{p^\rho}{p^2 -M^2} +
\frac{1}{1-a^2 p^2}\bigl ( a^2 \epsilon_{\mu\nu\rho}{p^\rho} + i a 
{\cal T}_{\mu\nu}),\label{comentario1}
\end{equation}

\noindent
where $M$ is the auxiliary massive parameter as introduced in Eq. (\ref{31})
and ${\cal T}_{\mu\nu}$ is the transversal projector defined below
Eq. (\ref{4C3}).  Observe that $a$ has dimension $-1$ in units of mass.  The
degree of superficial divergence is now given by

\begin{equation}
d(\Gamma)=
3- \frac12 N_A- \frac12 N_\phi-V_{\phi^4}- V_{A^2\phi^2}- \frac12 V_{A\phi^{*}
\partial \phi}\label{comentario}
\end{equation}

\noindent
where $V_{\cal O}$ denotes the number of vertices associated to ${\cal O}$ in
$\Gamma$.  With this new power counting many graphs previously
divergent turn out to be convergent.  In this situation, the following
possibilities can be envisaged.

i) Instead of (\ref{2}) one adopts (\ref{comentario}) to define the graphs
to be subtracted. The outcome is a well defined theory as far as $a$,
the coefficient of the Maxwell term, is kept nonvanishing.  The result
however is not analytic in $a$ and the limit $a\rightarrow 0$ does not
exist.

ii) One uses the old power counting and the parameter $a$ is not changed
in the subtraction terms.  This means that many graphs will be
oversubtracted. However such subtractions are needed if, as
$a\rightarrow 0$, one wants to recover the pure CS model
studied in this paper.

Finite one-loop renormalization constants for the non-Abelian CS
theory with fermionic matter were computed in \cite{chaichian} using
{\it consistent dimensional renormalization} and found to be different
from the ones obtained from the {\it dimensional reduction}
prescription. It should be noticed that, as remarked in \cite{asorey},
even for finite theories there could exist different families of BRST
invariant regularizations leading to distinct results for the one-loop
radiative corrections. This may imply that in spite of the numerical
differences there are no physical inconsistencies between the two
approaches.

The model studied in \cite{pinheiro} is a particular case of Eq.
(\ref{1}), with $\lambda=m^2=0$. We found a complete agreement in
$\gamma_{\Phi}^{[2]}$, $\beta_\nu^{[2]}$, and 
$\gamma_{\Phi^\dagger \Phi}^{[2]}$.  Our
results show that the dimensional reduction method is consistent at
two-loops, at least, as far as the computation of the RG functions is
concerned.

\begin{center}
{ACKNOWLEDGMENTS}
\end{center}

 This work was partially supported by Funda\c c\~ao de Amparo
\`a Pesquisa do Estado de S\~ao Paulo (FAPESP) and Conselho Nacional de
Desenvolvimento Cient\'\i fico e Tecnol\'ogico (CNPq).
L. C. A. would like to thank the Mathematical Physics Department for
 their hospitality.

\appendix
\section{}

\begin{figure}
\centerline{ \epsfbox{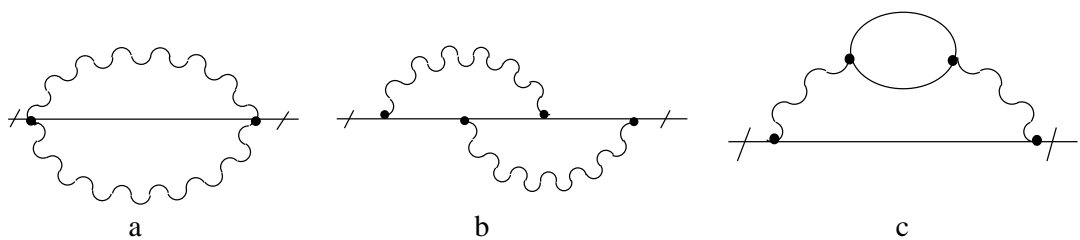}} \caption{{\it Contributions to $\gamma_\Phi$.}} \label{fig4}
\end{figure}

\begin{figure}
\centerline{ \epsfbox{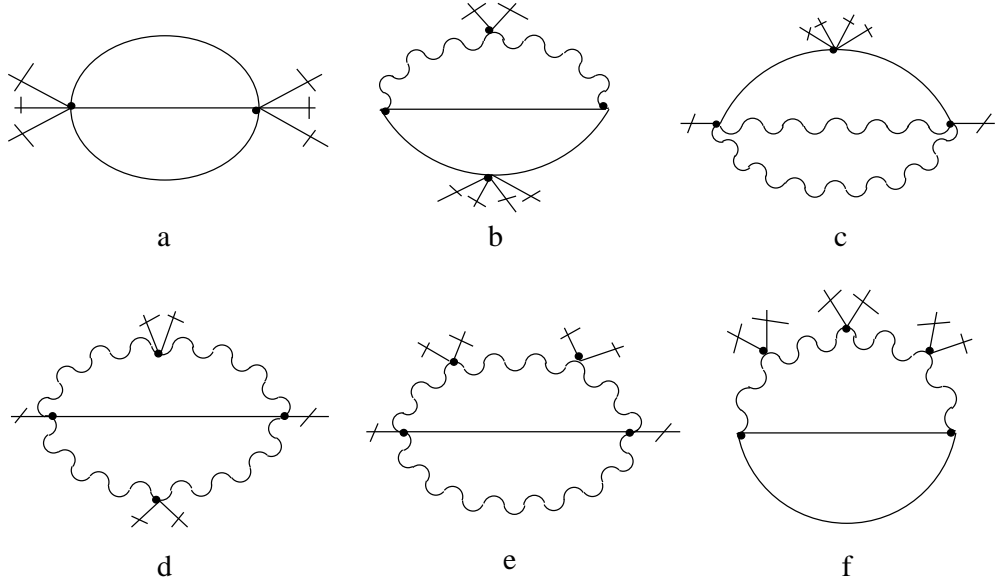}} \caption{{\it Two-loop contributions to $\Gamma^{(6)}$.}} \label{fig5}
\end{figure}

\begin{figure}
\centerline{ \epsfbox{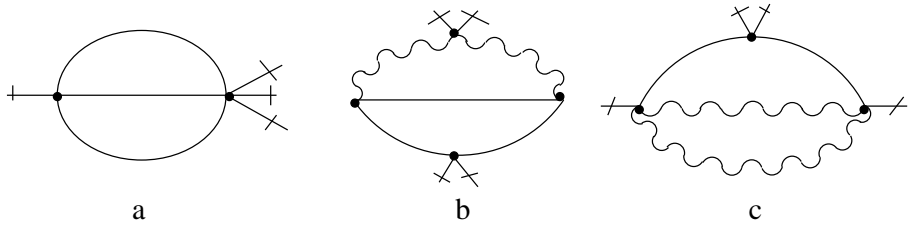}} \caption{{\it Contributions to $\Gamma^{(4)}$ at two loop.}} \label{fig6}
\end{figure}

\begin{figure}
  \centerline{ \epsfbox{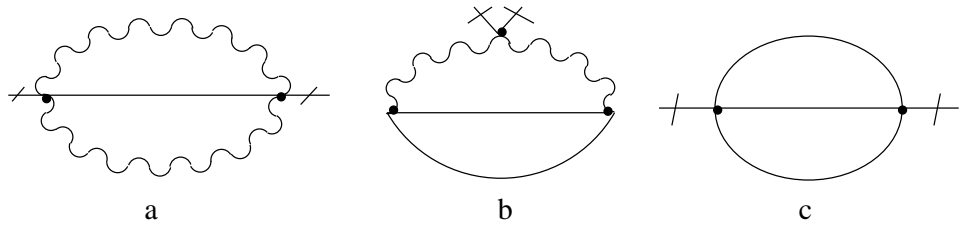}} \caption{{\it Contributions to $\Sigma^{[2]}(p=0)$. }} \label{fig7}
\end{figure}

In this appendix we shall present details of the calculation of the
contribution of the diagram \ref{fig4}c to $\gamma^{[2]}_{\Phi}$. The graph
($\Gamma$) is logarithmically divergent and has just one divergent
subgraph which is the same as the one associated to the $\pi^{B}_{\rho\lambda}$
contribution to the CS vacuum polarization. We will denote this subgraph
by $\gamma$. There are four forests: $\emptyset$ (the empty forest), $\Gamma$,
$\gamma$ and $\{\Gamma, \, \gamma\}$. Thus, the application of the {\it forest
formula} to (\ref{26}) leads to

\begin{eqnarray}
&&\Sigma_{C}^{[2]}(p)=16 i e^4 p^\mu\, p^\nu \int [dk][dq] \Bigl \{
q^\alpha q^\beta \Delta_{\mu\alpha}(k) \Delta_{\beta \nu}(k)\Bigl [\Delta(k-p,m)\Delta(q,m)\Delta (q-k,m)\Bigr.\Bigr.\nonumber \\
&&\Bigl.-\Delta(k,\mu)\Delta(q,\mu)\Delta (q-k,\mu)\Bigr ]-q^\alpha q^\beta \Delta_{\mu\alpha}(k) \Delta_{\beta \nu}(k)\Delta(k-p,m)\Bigl [\Delta^2(q,0)- i k\cdot q \Delta^3(q,\mu)\Bigr ]\nonumber\\
&&+q^\alpha q^\beta \Delta_{\mu\alpha}(k) \Delta_{\beta \nu}(k)\Delta(k,\mu))
\Bigl [\Delta^2(q,0)-ik\cdot q \Delta^3(q,\mu)\Bigr ]
\Bigr \}
\end{eqnarray}

\noindent
The divergent parts of the terms containing the $k\cdot q$ factor cancel between themselves; the finite parts associated to them are odd and vanish under
symmetrical integration. We can therefore rewrite the above expression as

\begin{eqnarray}
&&\Sigma_{C}^{[2]}(p)=16  e^4 p^\mu\, p^\nu \epsilon_{\mu\alpha\lambda}
\epsilon_{\beta \nu\sigma}\int [dk][dq]k^\lambda k^\sigma\Biggl \{
\frac1{[(k-p)^2-m^2][q^2-m^2][(q-k)^2-m^2]}\Bigr .\nonumber\\
&&\Bigl.-\frac1{[k^2-\mu^2][q^2-\mu^2][(q-k)^2-\mu^2]}-
\frac1{[(k-p)^2-m^2](q^2)^2}+\frac1{[(k^2-\mu^2](q^2)^2}\Biggr \}\, .
\end{eqnarray}

The contribution to $\gamma^{[2]}_{\Phi}$ arising from the above integral
can be most easily computed if one employs an intermediate auxiliary
regularization. Adopting dimensional regularization  we arrive at
\begin{equation} 
\gamma^{[2]}_{\Phi C}=-{16e^4 }[X_1-X_2],
\end{equation}

\noindent
where
\begin{eqnarray}
X_1&=&\int [dk][dq]\mu\frac{\partial\phantom {a}}{\partial\mu}\frac{q^2}
{k^2(k^2\mu^2)(q^2-\mu^2)((k+q)^2-\mu^2)},\\
X_2&=&\int [dk][dq]\mu\frac{\partial\phantom {a}}{\partial\mu}\frac{(k\cdot q)^2}
{(k^2)^2(k^2-\mu^2)(q^2-\mu^2)((k+q)^2-\mu^2)}
\end{eqnarray}

\noindent
which, after some simple manipulations, can be calculated with the help of
the results (\ref{22}) and (\ref{23}). We obtain
\begin{eqnarray}
X_1(\mu) & = & \frac{1}{960\pi^2},\\
X_2(\mu) & = & -\frac{1}{640\pi^2}\, ,
\end{eqnarray}

from which we obtain the result quoted in the text.

\begin{table}
\caption{ Contributions to $\Gamma^{(6)[2]}$ arising from the graphs in
Fig. \ref{fig5}. \label{tb1}}
\begin{tabular}{cc}
{Diagram type}& {Contribution}\\
\tableline
5a&$-\frac{7i}{24\pi^2}\,\nu^2\,\ln\,\left(\frac{m}{\mu}\right )$\\
5b&$\frac{9i}{4\pi^2}\,e^4\nu\,\ln\left (\frac{m}{\mu}\right)$\\
5c&$\frac{15i}{4\pi^2}\,e^4\nu\,\ln\left(\frac{m}{\mu}\right)$\\
5d&$-\frac{27i}{\pi^2}\,e^8\,\ln\left(\frac{m}{\mu}\right)$\\
5e&$-\frac{9i}{\pi^2}\,e^8\,\,\ln\left(\frac{m}{\mu}\right)$\\
5f&$-\frac{9i}{\pi^2}\,e^8\,\,\ln\left(\frac{m}{\mu}\right)$\\
\end{tabular}
\end{table}


\end{document}